%Paper: cond-mat/9501020
%From: "Paul Krapivsky (617) 353-9885; FAX -9393" <paulk@buphy.bu.edu>
%Date: Fri, 6 Jan 1995 15:48:03 -0500 (EST)

%Comments: Plain TeX

\def\cp #1 #2 #3 {{\sl Chem.\ Phys.} {\bf #1}, #2 (#3)}
\def\jetp #1 #2 #3 {{\sl Sov.\ Phys.\ JETP} {\bf #1}, #2 (#3)}
\def\jpa #1 #2 #3 {{\sl J. Phys.\ A} {\bf #1}, #2 (#3)}
\def\jcp #1 #2 #3 {{\sl J.\ Chem.\ Phys.} {\bf #1}, #2 (#3)}
\def\jpc #1 #2 #3 {{\sl J.\ Phys.\ Chem.} {\bf #1}, #2 (#3)}
\def\jsp #1 #2 #3 {{\sl J.\ Stat.\ Phys.} {\bf #1}, #2 (#3)}
\def\jdep #1 #2 #3 {{\sl J.\ de Physique I} {\bf #1}, #2 (#3)}
\def\mclc #1 #2 #3 {{\sl Mol.\ Cryst.\ Liq.\ Cryst.} {\bf #1}, #2 (#3)}
\def\nat #1 #2 #3 {{\sl Nature} {\bf #1}, #2 (#3)}
\def\pA #1 #2 #3 {{\sl Physica\ A} {\bf #1}, #2 (#3)}
\def\pra #1 #2 #3 {{\sl Phys.\ Rev.\ A} {\bf #1}, #2 (#3)}
\def\prb #1 #2 #3 {{\sl Phys.\ Rev.\ B} {\bf #1}, #2 (#3)}
\def\pre #1 #2 #3 {{\sl Phys.\ Rev.\ E} {\bf #1}, #2 (#3)}
\def\prl #1 #2 #3 {{\sl Phys.\ Rev.\ Lett.} {\bf #1}, #2 (#3)}
\def\rmp #1 #2 #3 {{\sl Rev.\ Mod.\ Phys.} {\bf #1}, #2 (#3)}
\def\zpc #1 #2 #3 {{\sl Z. Phys.\ Chem.} {\bf #1}, #2 (#3)}
\def\zw #1 #2 #3 {{\sl Z. Wahrsch.\ verw.\ Gebiete} {\bf #1}, #2 (#3)}

%%%%%%%%%%%%%%%%%%%%%%%%%%%%%%%%%%%%%%%%%%%%%%%%%%%%%%%%%%%%%%%%%%%%%%%%%%%%%%
\def\gtwid{\mathrel{\raise.3ex\hbox{$>$\kern-.75em\lower1ex\hbox{$\sim$}}}}
\def\ltwid{\mathrel{\raise.3ex\hbox{$<$\kern-.75em\lower1ex\hbox{$\sim$}}}}
\def\eg{{\it e.\ g.}}\def\ie{{\it i.\ e.}}
\def\ea{{\it et al.}}

\def\pd#1#2{{\partial #1\over\partial #2}}      %partial derivative
\def\p2d#1#2{{\partial^2 #1\over\partial #2^2}} %second partial derivative
      %total derivative
\def\t2d#1#2{{d^2 #1\over d #2^2}} %second total derivative

\newcount\refnum\refnum=0  %automatic, non-labelled references
\def\refi{\smallskip\global\advance\refnum by 1\item{\the\refnum.}}
\newcount\rfignum\rfignum=0  %automatic, non-labelled figure labels in captions
\def\rfigi{\medskip\global\advance\rfignum by 1\item{Figure \the\rfignum.}}

\newcount\eqnum \eqnum=0  %automatic, non-labelled equation numbering
\def\eqnoi{\global\advance\eqnum by 1\eqno(\the\eqnum)}
\def\eqnai{\global\advance\eqnum by 1\eqno(\the\eqnum {\rm a})}
\def\eqnbi{\eqno(\the\eqnum {\rm b})}
\def\back#1{{\advance\eqnum by-#1 Eq.~(\the\eqnum)}}%refer to one equation back
\def\last{Eq.~(\the\eqnum)}                   %refer to the current equation

%%%%%%%%%%%%%%%%%%%%%%%%%%%%%%%%%%%%%%%%%%%%%%%%%%%%%%%%%%%%%%%%%%%%%%%%%%%%%%

\overfullrule=0pt
\magnification 1200
\baselineskip=20 true bp

%%%%%%%%%%%%%%%%%%%%%%%%%%%%%%%%%%%%%%%%%%%%%%%%%%%%%%%%%%%%%%%%%%%%%%%%%%%%

\def\a{\alpha}
\def\b{\beta}
\def\g{\gamma}

\def\m{\mu}
\def\n{\nu}
\def\l{\lambda}
\def\t{\tau}

\def\Caxt{{\cal A}(xt^{-\alpha})}
\def\Cbxt{{\cal B}(xt^{-\alpha})}
\def\Crxt{{\cal R}(xt^{-\alpha})}

\def\Cax{{\cal A}(\xi)}
\def\Cbx{{\cal B}(\xi)}

\def\Cb{{\cal B}}
\def\Crx{{\cal R}(\xi)}

\def\arrt{a(x+{\bf r},t+\t)}
\def\axt{a(x,t)}
\def\bxt{b(x,t)}
\def\rxt{r(x,t)}
\def\coef{\left({c_0^2\over Dt}\right)^{1\over 4}}
\def\coeff{\left({c_0^2\over Dt}\right)^{3\over 4}}
\def\coefmf{\left({c_0^2\over Dt}\right)^{1\over 3}L^{\mu}}
\def\coeffmf{\left({c_0^2\over Dt}\right)^{2\over 3}L^{\nu}}

\def\lap{2D\Delta_{\bf r}}
\def\dx{\Delta x}
\def\erf{c_0~{\rm erf}\left({x\over \sqrt{4Dt}}\right)}

%%%%%%%%%%%%%%%%%%%%%%%%%%%%%%%%%%%%%%%%%%%%%%%%%%%%%%%%%%%%%%%%%%%%%%%%%%%%%%

\centerline{\bf Diffusion-Limited Annihilation with Initially
Separated Reactants}
\vskip 0.4in
\centerline{P.~L.~Krapivsky}
\bigskip
\centerline{\sl Center for Polymer Studies and Department of Physics}
\centerline{\sl Boston University, Boston, MA 02215}
\vskip 0.6in
\centerline{ABSTRACT}\smallskip
{\narrower\narrower\noindent
A diffusion-limited annihilation process, $A+B \to 0$,
with species initially separated in space is investigated.
A heuristic argument suggests the form of the reaction rate in
dimensions less or equal to the upper critical dimension $d_c=2$.
Using this reaction rate we find that the width of the reaction
front grows as $t^{1/4}$ in one dimension and as
$t^{1/6}(\ln t)^{1/3}$ in two dimensions.\bigskip}

\noindent
P. A. C. S. Numbers: 82.20.-w, 05.40.+j., 02.50.-r, 82.70.-y.

\vfil\eject
%\vskip 0.8in

%%%%%%%%%%%%%%%%%%%%%%%%%%%%%%%%%%%%%%%%%%%%%%%%%%%%%%%%%%%%%%%%%%%%%%%%%%%

In this paper, we investigate the kinetics of diffusion-limited
two-species annihilation process, $A+B \to 0$, in which collisions
between two distinct species $A$ and $B$ lead to the formation
of inert reaction products. Classically, such a process can be
described by reaction-diffusion equations [1],
$$
\pd a t = D\p2d a x - r, \quad
\pd b t = D\p2d b x - r.
\eqnoi
$$
Here $a=\axt$ and $b=\bxt$ are the local concentrations of
$A$- and $B$-species, respectively, $D$ is the diffusion
coefficient which is assumed to be identical for both species,
and $r=\rxt$ the reaction rate. In the mean-field approximation
this reaction rate is proportional to concentrations of reactants,
$r=kab$. For sufficiently low dimensions, the diffusion mechanism
is not efficient enough and fluctuations in the densities of
diffusing reactants will result in a different form for the
reaction rate and dimension-dependent kinetic behavior at
long times [2].

Several recent studies have been focussed on the situation
in which species are initially separated in space and hence
they react in a confined region called the reaction front.
This front plays an important role for a variety of physical
and chemical processes [3]. At the  mean-field level,
a scaling theory has been developed [4] which predicts that
the width of the reaction front, $w$, grows in time as $t^{1/6}$
in agreement with experiments [5] and simulations [5-7].

Since the upper critical dimension for this process appears
to be $d_c=2$ [7], a departure from the value $\a={1\over 6}$
for the width exponent is expected for $d \le 2$. Numerical
simulations of one dimensional systems suggest the value
$\a \approx 0.3$ [7-9]. However, an exact value of the width
exponent in one dimension is unknown. In two recent works it
was argued that $\a={1\over 4}$ in one dimension. Both these
studies are based on analysis of more tractable systems:
the first paper [10] examines the case in which one reactant
is static and the other diffuses while the second paper [11]
treats the behavior of the front in the steady state reached
by imposing the antiparallel current densities of $A$ and $B$
species at $x=+\infty$ and $x=-\infty$, respectively.
The latter steady state version of the original model was
introduced in [12] and further explored in [13].

In the present study, we investigate the original transient
problem of initially separated equally mobile species.
Our analysis is based on the rate equations approach but with a
modified reaction rate. In one dimension, we find $\a={1\over 4}$
in agreement with previous studies. In two dimensions we derive
a logarithmic correction to the mean-field behavior:
$w \sim t^{1/6}(\ln t)^{1/3}$. While it may be difficult to
confirm this logarithmic correction numerically, a significant
departure from the mean-field behavior, $w \sim t^{1/6}$,
has been observed [7].

Let us first estimate the reaction constant $k$ in the mean-field
relation $r=kab$. It is reasonable to assume that the reaction
constant depends on the diffusion constant $D$ and the radius
of particles $R$ which, for simplicity, is assumed to be the
same for both species. Simple dimensional analysis gives
$k \sim DR^{d-2}$. On physical grounds, the reaction constant
must be an increasing function of the radius $R$ and hence the
previous expression for $k$ is valid only for $d>2$.
To find the reaction constant for $d<2$ we should assume
that $k$ does not depend on $R$ but instead it depends on
concentrations, $k=k(D,a,b)$. A dimensional analysis gives
$k\sim D(a+b)^{-1+2/d}$ and therefore
$$
r \sim Dab(a+b)^{-1+2/d}.
\eqnoi
$$
In fact, any symmetric function $r=r(a,b)$ of densities $a$ and
$b$ with degree of homogeneity $1+2/d$ is equally possible.
However, in finding scaling exponents the precise form of the
reaction term is unimportant and for concreteness we choose the
form given above. Note that this simple dimensional argument also
correctly predicts the value of the upper critical dimension, $d_c=2$.

This type of argument was first applied by Toussaint and Wilczek
[14] to homogeneous single-species annihilation, $A+A \to 0$,
and led to the qualitatively correct long-time kinetic behavior:
$a \sim (Dt)^{-d/2}$ for $d<2$. In contrast, for homogeneous
two-species annihilation, $A+B \to 0$, this argument predicts
the incorrect value of the upper critical dimension
(the correct value is $d_c=4$) and the incorrect kinetic behavior
($t^{-d/2}$ instead of $t^{-d/4}$, see [2,14]). Because of this
difficulty there were no attempts to apply the argument to
inhomogeneous two-species annihilation. It can be readily seen
that a large-scale spatial organization of reactants [2,14] for
homogeneous two-species annihilation leads to this difficulty.
Indeed, the reaction takes place only near boundaries between
different-species domains. The fraction of space occupied
by these reaction zones tends to zero as $t\to \infty$ [15].
Hence modeling the reaction term by a function of {\it averaged}
concentrations $a(t)$ and $b(t)$ is inappropriate since it
gives the same reaction rate inside single-species domains
where reaction does not proceed and in the boundaries. However,
for {\it inhomogeneous} two-species annihilation, the behavior
in reaction zone where the reaction does proceed is of interest.
Therefore, a description of reaction zone by using a (modified)
reaction rate which is a function of local concentrations $\axt$
and $\bxt$ seems reasonable.

Consider now a one dimensional inhomogeneous system in which
$A$ particles are uniformly distributed to the right of the
origin and $B$ particles are uniformly distributed to the
left of the origin, both with equal concentration $c_0$. The
difference $\axt-\bxt$ satisfies the diffusion equation which,
subject to given initial conditions, has the solution
$$
\axt-\bxt = \erf.
\eqnoi
$$
In the long-time limit, the scaling form
$$
\axt \sim t^{-\g}\Caxt, \quad
\bxt \sim t^{-\g}\Cbxt, \quad
\rxt \sim t^{-\b}\Crxt,
\eqnoi
$$
is expected. Since $r \sim (a+b)ab$ in one dimension (see \back2),
the scaling form of \last~ implies the scaling relation $\b=3\g$.
Further, by inserting \last~ into \back2~ and asymptotically balancing
various terms one finds another scaling relation $\g+2\a=3\g$.
Finally, assuming the reaction zone is short compared  to the
diffusion length $\sqrt{Dt}$ and making
use of the exact solution (3) one gets
$a-b \simeq c_0 x/\sqrt{\pi Dt} \sim t^{\a-1/2}$.
This gives the last scaling relation $\a+\g={1\over 2}$.
Combining these scaling relations we find the exponents:
$\a=\g={1\over 4}, \b={3\over 4}$. Note that the width
indeed increases slower than the diffusion length,
$w \sim t^{1/4} \ll  t^{1/2}$, thus confirming our assumption.
Rewriting the scaling ansatz in dimensionless form, we arrive at
$$
\axt \simeq \coef\Cax, \quad
\bxt \simeq \coef\Cbx, \quad
\rxt \simeq D\coeff\Crx,
\eqnoi
$$
with the scaling variable $\xi = x\left({c_0^2/Dt}\right)^{1/4}$.

Substituting Eqs.~(5) into the governing equations (1) and making
use of \back2~ we arrive at still intractable nonlinear ordinary
differential equation. However, if we focus on the region far
enough from the origin, $\xi \gg 1$, some progress can be made.
To find the scaling form of densities we have to choose an
appropriate form of the reaction rate. Let us mention that
the reaction rate given by \back3~ at $d=1$, $r \sim Dab(a+b)$,
has an additional advantage of being linear in concentration
which is small compared  to the other, \eg, $r\sim Da^2b\propto b$
if $b \ll a$. This physically appealing behavior suggests to use
this particular form.

So, let us examine Eqs.~(1) with the reaction rate
$$
r=\kappa Dab(a+b),
\eqnoi
$$
where $\kappa$ is a dimensionless constant. In the region
$\left({Dt/c_0^2}\right)^{1/4} \ll x \ll \sqrt{Dt}$ where
the $A$-species dominates \back3~ simplifies to
$a \simeq c_0 x/\sqrt{\pi Dt}$, which can be rewritten as
$\Cax \simeq \xi/\sqrt{\pi}$ in the scaling form.
Then the second Eq.~(1) asymptotically reduces to
$$
{d^2 \Cb\over d \xi^2} = {\kappa \over \pi} \xi^2\Cb.
\eqnoi
$$
By applying the WKB procedure [16] one finds the asymptotic
solution to \last,
$$
\Cbx \sim {1\over\sqrt{\xi}}\exp\left[-\sqrt{\kappa \over 4\pi}\xi^2\right],
\eqnoi
$$
which is valid for $\xi \gg 1$, \ie,
for $x \gg \left({Dt/c_0^2}\right)^{1/4}$.
Combining Eqs.~(5), (6), and (8) we arrive at the
final asymptotic expression for the reaction rate:
$$
\rxt \sim D\left({c_0^2\over Dt}\right)^{9/8}x^{3/2}
\exp\left[-\sqrt{\kappa \over 4\pi}~x^2
\left({c_0^2\over Dt}\right)^{1/2}\right],
\eqnoi
$$
for $\left({Dt/c_0^2}\right)^{1/4} \ll x \ll \sqrt{Dt}$.

Let us compare our prediction for the tail of the scaling
distribution, \last, with other studies. In a recent paper [8],
the validity of the scaling distribution has been challenged
and a description based on a continuous spectrum of scales
between $t^{1/4}$ and $t^{3/8}$ has been proposed. In particular,
Araujo \ea~ have fitted the tail of the reaction front by
the exponential form: $r \sim \exp\left[-|x|/ t^{3/8}\right]$.
In a very recent study [17], Cornell has reconsidered the problem.
Based on extensive simulations he has found that the dynamical
scaling appears to hold, contrary to the claim of Ref.~[8].
He has obtained $\a \approx 0.28$ for the width exponent
and observed that the reaction rate profile has a Gaussian form.
Thus our results well agree with the most extensive present
simulations [17].

Let us now turn to the two dimensional version of the problem.
The same line of reasoning gives the reaction rate
$r \sim Dab$ and hence the mean-field behavior could be
expected. However, $d=2$ is the marginal dimension, logarithmic
corrections to mean-field expression $r \sim Dab$ may arise,
and therefore one has to be more careful. To assess the validity
of the mean-field expression for the reaction rate let us start
from more fundamental relation, $r \sim b/T$, where $T$ is a time
interval in which a $B$-particle typically collides with some
$A$-particle. Consider for simplicity the region in which
$A$ species dominates. In a reference frame at rest with respect
to an arbitrary ``target'' $B$-particle, the density $\arrt$ of
$A$-particles is governed by the diffusion equation
$$
\pd{\arrt}{\t}=\lap \arrt.
\eqnoi
$$
We should solve this equation subject to the initial condition
$a(\t=0)=\axt$ (since the local density is $\axt$) and the adsorbing
boundary condition $a(|{\bf r}|=2R)=0$ for $\t > 0$.
When $D\t \gg R^2$, an approximate solution to \last~ with
prescribed initial and boundary conditions can be readily
found by using a quasistatic approximation (see, \eg, [18]).
This solution reads
$$
\arrt \simeq \axt{\ln(r/2R)\over 2\ln\left(D\t/R^2\right)}.
\eqnoi
$$
Note that the ``local'' density given by \last~ is changed
on the length scale of the order $\sqrt{D\t}$ which has to be small
compared to the length scale of the ``external'' density $\axt$,
\ie, $\sqrt{D\t} \ll w$. We will assume that both inequalities
$R^2 \ll \sqrt{DT} \ll w$ hold and therefore \last~ may be
applied to estimate the collision time $T$. From the final
expressions for $T$ and $w$ it is straightforward to verify
that previous inequalities are indeed satisfied in the long-time
limit thus providing the check of self-consistency.

The collision time $T$ may be evaluated by computing
the flux to the circle of radius $2R$ and then by equating
the flux to the unity:
$$
8\pi DR\int\limits_0^T\pd {a(|{\bf r}|=2R,t+\t)}{r} d\t = 1.
\eqnoi
$$
By combining \last~ and \back1~ one gets the final estimate,
$T \simeq \ln\left(1/2\pi aR^2 \right)/2\pi Da$. Thus we obtain
the following expression for the reaction rate,
$$
r=2\pi Dab\left[{1\over \ln\left(1/2\pi aR^2\right)}
+ {1\over \ln\left(1/2\pi bR^2\right)}\right].
\eqnoi
$$
The second term is added to keep $r=r(a,b)$ symmetric.
In the region $x \gg w$ which we have considered the second term
is small and thus never appeared. Similarly, in the region
$-x \ll -w$ the first term is negligible. Since the reaction
rate must be symmetric function of the densities, \last~ provides
a proper combine approximation in both tail regions.
It is therefore reasonable to assume that \last~ is in fact
accurate everywhere in reaction zone. So instead of the
mean-field expression for the reaction rate in two dimensions,
$r=Dab$, a refined argument gives a logarithmically corrected
value (13).

Now let us try a scaling ansatz
$$
\eqalign{\axt \simeq \coefmf\Cax, \quad
\bxt \simeq \coefmf\Cbx,\cr
\rxt \simeq D\coeffmf\Crx, \quad
\xi=x\left({c_0^2\over Dt}\right)^{1 \over 6}L^{\l},\cr}
\eqnoi
$$
where we have used a shorthand notation $L$ for the logarithmic
factor $L=\ln\left({Dt/c_0^2R^6}\right)$. \last~ is just
the standard mean-field scaling ansatz [4] modified by
logarithmic corrections. A procedure identical to the one used
in the one dimensional situation gives three scaling relations
between ``logarithmic'' exponents $\l,\m,$ and $\n$ from which
one finds: $\l=\n=-{1\over 3}, \m={1\over 3}$.

The main lesson from this study is the following:
Even for sufficiently low dimensions, $d \le d_c$,
when the strong correlations destroy the mean-field approximation,
it is still possible to use mean-field like diffusion-reaction
equations with modified reaction rates. At the critical dimension,
the present heuristic method gives subtle logarithmic corrections
which seem hard to obtain by more solid theoretical approaches.

Let us finally mention that the more general higher order
annihilation reaction process, $mA+nB \to 0$, with initially
separated species can be analogously investigated by the present
approach. Since the critical dimension for this reaction process
is $d_c={2 \over m+n-1}$ (see, \eg, [11]), deviations from the mean-field
predictions are expected only for the classic case $(m,n)=(1,1)$
[when $d=1$ and 2] and for $(m,n)=(2,1)$ [when $d=d_c=1$].
Consider the latter case and assume that point-like particles
undergo a random walk on the linear lattice with spacing $\dx$.
[Even in one dimension we must keep the lattice spacing finite
since otherwise the reaction with three-particle interaction
is absent.] Focus again on the region where $A$-species
dominates and estimate the collision time $T$. Note that the
collision event, \ie, the situation when two $A$-particles will
be simultaneously in the site with a target $B$-particle,
may be described as usual two-body collision between
an immobile target particle and imaginary particles
corresponding to {\it pairs} of $A$-particles and diffusing
on the two-dimensional square lattice with the same spacing $\dx$.
A more detailed description of this correspondence is given
in Ref.~[19] which considers a related single-species
aggregation process, $A+A+A \to A$. Then up to the replacement
of the one-dimensional density $a$ by the two-dimensional density
of pairs $a^2$ and the spacing $\dx$ by the radius $R$ one can
use previous results (see \back4-\back2) and find the collision
time: $T \simeq \ln\left(1/2\pi a^2\dx^2\right)/2\pi Da^2$.
So we obtain $r \sim Da^2b/\ln(1/a\dx)$ while the mean-field
approach would give $r \sim Da^2b$. This results in logarithmic
corrections to the mean-field behavior. Repeating the steps
employed for the previous treatment of two-particle annihilation
we arrive at the scaling form
$$
\axt \sim (L/t)^{1/4}\Cax, \quad
\bxt \sim (L/t)^{1/4}\Cbx, \quad
\rxt \sim t^{-3/4}L^{-1/4}\Crx.
\eqnoi
$$
Here $\xi=x\left(c_0^2/DtL\right)^{1/4}$ is the scaling
variable and $L=\ln\left({Dt/c_0^2 \dx^4}\right)$ the
logarithmic factor. Thus we conclude that the width of the
reaction front for the three-particle reaction process,
$2A+B \to 0$, scales as $t^{1/4}(\ln t)^{1/4}$ in one dimension.

\bigskip\bigskip

I am thankful to S.~Redner for discussions and a critical
reading of the manuscript and to E.~Ben-Naim for correspondence.
This work has been partially supported by grants from ARO and NSF.

\vfil\eject

\centerline{\bf References}\smallskip

\refi S.~A.~Rice, {\sl Diffusion-Controlled Reactions} (Elsevier,
      Amsterdam, 1985).
\refi K.~Kang and S.~Redner, \pra 32 435 1985 .
\refi D.~Avnir and M.~Kagan, \nat 307 717 1984 ;
      G.~T.~Dee, \prl 57 275 1986 ;
      H.~K.~Henisch, {\sl Periodic Precipitation} (Pergamon, Oxford, 1991).
\refi L.~G\'alfi and Z.~R\'acz, \pra 38 3151 1988 .
\refi Y.~E.~Koo, L.~Li, and R.~Kopelman, \mclc 183 187 1990 ;
      Y.~E.~Koo and R.~Kopelman, \jsp 65 893 1991 .
\refi Z.~Jiang and C.~Ebner, \pra 42 7483 1990 .
\refi S.~Cornell, M.~Droz, and B.~Chopard, \pra 44 4826 1991 .
\refi M.~Araujo, H.~Larralde, S.~Havlin,
      and H.~E.~Stanley, \prl 71 3824 1993 .
\refi H.~Larralde, M.~Araujo, S.~Havlin,
      and H.~E.~Stanley, \pra 46 855 1992 .
\refi H.~Larralde, M.~Araujo, S.~Havlin,
      and H.~E.~Stanley, \pra 46 6121 1992 .
\refi S.~Cornell and M.~Droz, \prl 70 3824 1993 .
\refi E.~Ben-Naim and S.~Redner, \jpa 25 L575 1992 .
\refi B.~Lee and J.~Cardy, \pre 50 3287 1994 .
\refi D.~Toussaint and F.~Wilczek, \jcp 78 2642 1983 .
\refi F.~Leyvraz and S.~Redner, \pra 46 3132 1992 .
\refi C.~M.~Bender and S.~A.~Orszag, {\sl Advanced Mathematical Methods
      for Scientists and Engineers} (McGraw-Hill, Singapore, 1984).
\refi S.~Cornell, unpublished (cond-mat/9410061).
\refi S.~Redner and D.~ben-Avraham, \jpa 23 L1169 1990 ;
      P.~L.~Krapivsky, \pre 47 1199 1993 .
\refi P.~L.~Krapivsky, \pre 49 3233 1994 .

\vfill\eject\bye

\end